\documentclass[aps,prl,twocolumn,groupedaddress,showpacs,nofootinbib]{revtex4}
\usepackage{amsmath} 
\usepackage{amssymb} 
\usepackage{amsxtra}
\usepackage{graphicx}
\usepackage{psfrag} 
\usepackage{bm} 
\usepackage{float} 
\usepackage{dsfont}
\usepackage{hyperref}
\usepackage{color}

\usepackage{ulem}
\definecolor{darkblue}{rgb}{0.0,0,0.6}

\begin{document}

\newcommand{\Sref}[1]{Section~\ref{#1}}
\newcommand{\sref}[1]{Sec.~\ref{#1}}
\newcommand{\Cref}[1]{Chap.~\ref{#1}}
\newcommand{\tql}{\textquotedblleft} 
\newcommand{\tqr}{\textquotedblright~} 
\newcommand{\tqrc}{\textquotedblright} 
\newcommand{\Refe}[1]{Equation~(\ref{#1})}
\newcommand{\Refes}[1]{Equations~(\ref{#1})}
\newcommand{\fref}[1]{Fig.~\ref{#1}}
\newcommand{\frefs}[1]{Figs.~\ref{#1}}
\newcommand{\Fref}[1]{Figure~\ref{#1}}
\newcommand{\Frefs}[1]{Figures~\ref{#1}}
\newcommand{\reff}[1]{(\ref{#1})}
\newcommand{\refe}[1]{Eq.~(\ref{#1})}
\newcommand{\refes}[1]{Eqs.~(\ref{#1})}
\newcommand{\refi}[1]{Ineq.~(\ref{#1})}
\newcommand{\refis}[1]{Ineqs.~(\ref{#1})}
\newcommand{\framem}[1]{\overline{\overline{\underline{\underline{#1}}}}}
\newcommand{\PRA }{{ Phys. Rev.} A }
\newcommand{\PRB }{{ Phys. Rev.} B} 
\newcommand{\PRE }{{ Phys. Rev.} E}
\newcommand{\PR}{{ Phys. Rev.}} 
\newcommand{\APL }{{ Appl. Phys. Lett.} }
\newcommand{\PRL}{Phys.\ Rev.\ Lett. }
\newcommand{\OCOM }{{ Opt. Commun.} } 
\newcommand{\JOSA }{{ J. Opt. Soc. Am.} A}
\newcommand{\JOSB }{{ J. Opt. Soc. Am.} A}
\newcommand{\JMO }{{J. Mod. Opt.}}
\newcommand{\RMP}{Rev. \ Mod. \ Phys. }
\newcommand{\etal} {{\em et al.}}
\newcommand{\etali} {{\em etal.}}

\accentedsymbol{\dbareps}{\Bar{\Bar{\epsilon}}}
\accentedsymbol{\dbarA}{\Bar{\Bar{A}}}
\accentedsymbol{\dbarB}{\Bar{\Bar{B}}}
\newcommand{\vv}{\mathbf}
\newcommand{\tens}[1]{\Bar{\Bar {#1}}}
\newcommand{\intover}{\mathrm d}
\newcommand{\dd}[2]{\frac{\intover {#1}}{\intover {#2}}}

\newcommand{\expval}[1]{\left\langle #1 \right\rangle} 
\newcommand{\abs}[1]{\left| #1 \right|}

\title{Few-photon transport in low-dimensional systems: Interaction-induced radiation trapping}

\author{Paolo Longo}
\affiliation{Institut f\"{u}r Theoretische Festk\"{o}rperphysik, 
             Universit\"{a}t Karlsruhe (TH), 76128 Karlsruhe, Germany}             
           
\author{Peter Schmitteckert}
\affiliation{Institut f\"{u}r Nanotechnologie, 
             Forschungszentrum Karlsruhe in der Helmholtz-Gemeinschaft,
             76021 Karlsruhe, Germany}

\author{Kurt Busch}
\affiliation{Institut f\"{u}r Theoretische Festk\"{o}rperphysik
             and DFG-Center for Functional Nanostructures (CFN), 
             Universit\"{a}t Karlsruhe (TH), 76128 Karlsruhe, Germany}

\date{\today}

\begin{abstract}
We present a detailed analysis of the dynamics of photon transport 
in waveguiding systems in the presence of a two-level system. 
In these systems, quantum interference effects generate a strong
effective optical nonlinearity on the few-photon level.
We clarify the relevant physical mechanisms through an appropriate
quantum  many-body approach. Based on this, we demonstrate that
a single-particle photon-atom bound state with an energy outside 
the band can be excited via multi-particle scattering processes.
We further show that these trapping effects are robust and, therefore, 
will be useful for the control of photon entanglement in solid-state 
based quantum optical systems.
\end{abstract}

\pacs{42.50.Nn, 42.50.-p, 42.50.Gy, 42.70.Qs}

\maketitle
Over the past years, the conception and development of solid-state 
based quantum optical functional elements have received steadily 
increasing interest \cite{Santori02,Politi08,Hofheinz08}.
As compared to other approaches, solid-state-based systems offer an 
obvious scalability and handling advantage of the resulting devices
as well as the utilization of modified light-matter interactions 
through judicious designs of the corresponding waveguides' dispersion 
relations and/or mode profiles.

However, since high-quality samples such as 
coupled-optical-resonator-waveguide arrays (CROWs) \cite{Xia07,Notomi08} 
have become available only recently, there is limited theoretical work 
regarding the potential of utilizing modified light-matter interaction 
in (effectively) low-dimensional quantum-optical systems. The basic underlying 
problem, i.e., that of a system with discrete levels that is coupled to a 
continuum of states has attracted attention for a long time \cite{Fano}.
For single 
photons, quantum interference effects in one-dimensional waveguides 
with an embedded quantum impurity allow the realization of effective 
energy-dependent mirrors \cite{Shen05,Zhou08,Longo09}. For two or more 
photons, this system induces an effective photon-photon interaction and 
even bound photon-photon states that may be exploited for efficient control 
of photon-entanglement \cite{Shen07a,Shen07b,Shi09}. Except for our work 
on the one-photon case \cite{Longo09}, all of the above calculations have 
been carried out in the stationary regime. 
In particular,
the more challenging few-photon case has been addressed with sophisticated 
Bethe-Ansatz \cite{Shen07a,Shen07b} and Lehmann-Symanzik-Zimmermann
reduction techniques \cite{Shi09} that allow one to determine the 
corresponding scattering matrices for such systems. However, these 
field-theoretical approaches employ linearized dispersion relations 
without band edges.  

In the present Letter, we apply our computational framework of 
time-domain simulations using Krylov-subspace-based operator-exponential 
methods \cite{Longo09,Schmitteckert04} to the case of few-photon transport through a quantum 
impurity in a one-dimensional waveguiding system 
similar to wave packet dynamics in electronic systems \cite{Schmitteckert04,Ulbricht09}.
This allows us to analyze the scattering of two or more photons at the quantum impurity 
in a very general way.
In particular, for a cosine-type dispersion relation, we are able to 
confirm the existence of two bound photon-atom states \cite{Shi09}. 
Furthermore, we show how these states can be excited and controlled 
through the photon-nonlinearity that is induced by the quantum impurity. 
This elucidates the mechanism through which the quantum impurity can 
be utilized for controlling photon-entanglement. In the field-theoretical
approaches discussed above \cite{Shen07a,Shen07b,Shi09}, the photon-atom
bound states are (due to the absence of band edges) energetically shifted 
to infinity and are thus removed from the physically accessible Hilbert 
space.

Starting from the well-known Dicke-Hamiltonian \cite{Dicke54}, we can derive 
a tight-binding 
Hamiltonian that describes photon propagation in an effectively one-dimensional 
waveguide with cosine-type dispersion relation (such as the CROWs of Refs. 
\cite{Xia07,Notomi08}) that is coupled to a quantum impurity as \cite{Longo09}
\begin{eqnarray}
\nonumber
\hat{H} & = & -J \sum_{x=1}^{N-1} \left( a_x^\dagger a_{x+1} + a_{x+1}^\dagger a_x \right) 
              + \frac{\Omega}{2} \sigma_z \\ 
\label{eq:TB-TLS}
        &   & + V \left( a_{x_0} \sigma_+ + a_{x_0}^\dagger \sigma_- \right).
\end{eqnarray}
Here, $a_x^\dagger$ and $a_x$ denote, respectively, bosonic (photon) creation
and annihilation operators at lattice site $x$ and $J$ denotes the corresponding
hopping element. The quantum impurity is modeled as a two-level system (TLS) 
with transition frequency $\omega_0 = \Omega/\hbar$ that is located at lattice 
site $x_0$ and couples with a coupling element $V$ to the modes of the photonic 
band. 
When measuring energies from the center of the band, the corresponding dispersion 
relation is $\hbar \omega_k = -2J \cos(ka)$, where $a$ denotes the lattice 
constant and $k$ stands for a wave number that lies within the first Brillouin 
zone. 
Finally, the TLS is described through the Pauli-operators $\sigma_z$ and 
$\sigma_\pm = \sigma_x \pm i \sigma_y$.

While being physically intuitive, the above Hamiltonian (\ref{eq:TB-TLS})  
does not allow for the most transparent discussion of the underlying
physics. Instead, we find it most useful to reformulate the problem in 
terms of the Hamiltonian
\begin{eqnarray}
\nonumber
\hat{H} & = & -J \sum_{x=1}^{N-1} \left( a_x^\dagger a_{x+1} + a_{x+1}^\dagger a_x \right)
            + \Omega b^\dagger b \\ 
\label{eq:TB-U}
        &   & + V \left( a_{x_0} b^\dagger + a_{x_0}^\dagger b \right) 
              + U b^\dagger b \left( b^\dagger b -1 \right), 
\end{eqnarray}
where we have replaced the TLS by an additional bosonic lattice site. More 
precisely, we have replaced the Pauli-operators of the TLS by appropriate 
combinations of bosonic creation and annihilation operators, $b^\dagger$ 
and $b$. The ground and excited states of the TLS correspond, respectively, 
to none and a single boson on this additional site (TLS site). Unphysical 
multiple occupancies of the TLS site have been addressed through the 
addition of the last term on the r.h.s. of (\ref{eq:TB-U}). This term 
ensures that once the TLS site is occupied, i.e., the TLS is in its excited 
state, adding a further boson to the TLS site requires the energy $U>0$. 
Thus, Hamiltonians (\ref{eq:TB-TLS}) and (\ref{eq:TB-U}) are equivalent in 
the limit $U \to \infty$ and this is the only case we consider in this work. 
The $U$-term induces inelastic scattering that allows us to discuss the 
physically relevant processes. For actual numerical calculations, we use
Hamiltonian (\ref{eq:TB-TLS}).

With this reformulation several issues become apparent. Quantum interference
processes associated with the coupling between TLS and the waveguide modes 
induce an effective interaction between photons as described by the nonlinear 
term $U b^\dagger b \left( b^\dagger b -1 \right)$. 
While this effective few-photon optical nonlinearity is spatially localized 
to the immediate vicinity of the TLS site, this system nevertheless represents 
a true quantum-mechanical many-particle problem. For instance, Hamiltonian
(\ref{eq:TB-U}) looks very similar to a bosonic version of the celebrated
single-impurity Anderson model \cite{Anderson61} that describes magnetic
impurities in metals.
Therefore, it is suggestive to apply methods that have been developed for
correlated quantum systems to the Hamiltonians (\ref{eq:TB-TLS}) and
(\ref{eq:TB-U}) \cite{Shen07a,Shen07b,Shi09,Longo09}.
From Hamiltonian (\ref{eq:TB-U}), it becomes apparent that the TLS will 
induce correlations between two or more photons. This raises the fascinating 
question to what extent the TLS can be utilized to engineer this entanglement 
and what role the photon-atom bound states play in this
(note that photon-atom bound states have been discussed in a different
context before \cite{Quang94}). 

\begin{figure}
\includegraphics[width=0.45\textwidth]{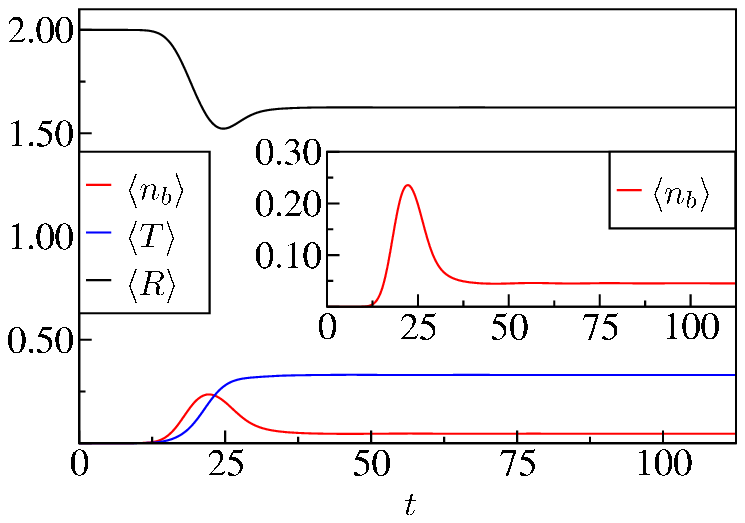}
\caption{\label{fig:bound-state-demo}
(Color online) 
Time evolution (in units of $\hbar / J$) of transmission 
$\langle T \rangle $, reflection $\langle R \rangle$, 
and impurity occupation 
$\langle n_b \rangle = \langle b^\dagger b \rangle = \langle \sigma_z +1/2 \rangle$ 
for a two-photon wave packet that scatters at a TLS. The TLS has a transition
energy $\Omega = \sqrt{2} J $ and couples with coupling strength $V = J$ 
to the central lattice site $x_0 = 100a$ of a tight-binding lattice with total
extent $L = 199a$ and hopping element $J$.
The photons are described via boson-symmetric wave packets that are constructed 
from single-particle Gaussian wave functions of width $s=6a$ with wave number 
$k = 3 \pi / 4a$ and initial center $x_c = 70a$ (see text and Ref. \cite{Longo09} 
for further details).
All calculations are stopped at times not exceeding the transit
time, i.e., the time the wave packet needs to pass through the waveguide,
thus avoiding artificial reflections from the system's boundaries.
}
\end{figure}

To address this question, we have to go beyond stationary calculations 
that determine the scattering matrices of photons in plane wave states 
for linearized dispersion relations where the photon-atom bound states
are physically inaccessible 
\cite{Shen07a,Shen07b,Shi09}. To do so, we employ our computational 
framework which we have described in detail elsewhere \cite{Longo09}.
This framework allows us to analyze both the dynamics of multi-photon 
wave packets that interact with the TLS and the dynamics of the TLS 
itself. Furthermore, it takes into account all aspects introduced by
the finite-bandwidth dispersion relation.
First, we would like to note that on energetic grounds a single photon
cannot excite the photon-atom bound states described above and, therefore, 
these states are of no relevance in single-photon scattering calculations 
from a TLS
\cite{Zhou08,Longo09}. 
In other words, the TLS (partially) absorbs an incoming single photon 
and a decomposition of the system's initial state into the (polaritonic) 
single-particle eigenstates of the Hamiltonian (\ref{eq:TB-TLS}) does 
not involve the bound photon-atom states. Thus, the excited TLS will 
eventually decay into its ground state. 
However, our reformulated Hamiltonian (\ref{eq:TB-U}) suggests that, by 
virtue of the nonlinear interaction term, the bound states can, in 
principle, be energetically reached via multi-photon processes.  
In Fig. \ref{fig:bound-state-demo}, we demonstrate that this is indeed 
possible: 
A two-photon wave packet interacts with the TLS and a sizable 
fraction of the photon population becomes trapped at the TLS site. 
In other words, once the TLS is appreciably excited by one of the incoming 
photons, the remaining photon sees a modified (saturated) TLS and is thus 
(partially) scattered into the hitherto unreachable bound photon-atom
states via multi-particle scattering processes. After the scattering
is complete, the bound photon-atom states are again decoupled from the 
continuum (such as is the case for the scattering of a single photon 
discussed above) and, thus, cannot decay. 
These bound states are of a polaritonic nature, i.e., 
they are multi-moded dressed eigenstates of (\ref{eq:TB-TLS}) 
and (\ref{eq:TB-U}) with complex wavenumbers solely induced by the 
existence of the waveguide's finite bandwidth.
This implies that a fraction of the radiation 
remains trapped at the TLS site in form of a partial occupation of 
the TLS.

\begin{figure}
\includegraphics[width=0.45\textwidth]{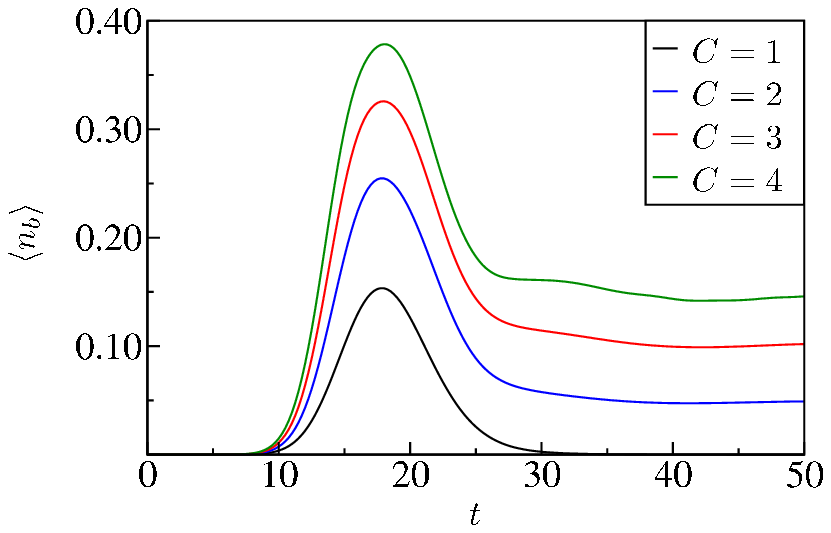}
\caption{\label{fig:multi-photon}
(Color online)
Time evolution (in units of $\hbar / J$) of the impurity occupation 
$\langle n_b \rangle$ for initial multi-photon states with different
photon numbers $C$ that are constructed analogous to the two-photon
states in Fig. \ref{fig:bound-state-demo}. The corresponding system
parameters are $L = 99a$, $x_0 = 50a$, $\Omega = \sqrt{2}J$, and 
$V = J$. The photon parameters are $x_c = 25a$, $s = 5a$, and
$k = 3 \pi / 4a$. The results for photon numbers $C=3$ and $C=4$
have been obtained with a time-dependent density matrix renormalization 
group (DMRG) technique as described in Ref. \cite{Schmitteckert04}.
}
\end{figure}

\begin{figure}
\includegraphics[width=0.45\textwidth]{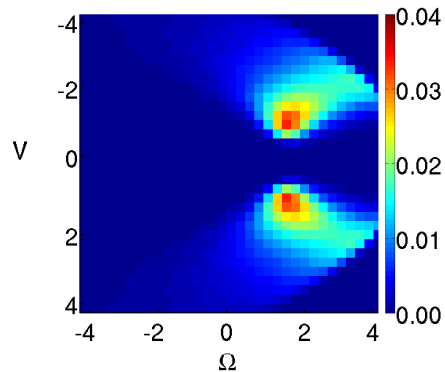}
\caption{\label{fig:detuning-coupling-3pi4}
(Color online)
Impurity occupation $\langle n_b \rangle$ in the long-time limit
(see Fig. \ref{fig:bound-state-demo}) after scattering of
two-photon states for different system parameters $V$ (in units
of $J$) and $\Omega$ (in units of $J$).
The fixed parameters are: $L = 199a$, $x_0 = 100a$, $x_c = 70a$, 
$s = 12a$, and $k = 3 \pi / 4a$.
}
\end{figure}

In order to verify the role of the multi-particle processes, we display
in Fig. \ref{fig:multi-photon} the time evolution of the TLS' 
excited-state occupation for the scattering
of multi-photon wave packets with different particle numbers.
The increase in the trapped photon population with the number
of photons implies a corresponding increase in the rate at which 
radiation is scattered into the bound states. The strength of this 
interaction further depends on the detuning of the TLS relative to the 
photon frequency as well as on the strength of the coupling matrix 
element $V$ between TLS and the waveguide modes. 
In Fig. \ref{fig:detuning-coupling-3pi4}, we depict the corresponding 
dependence of the trapped photon population at the TLS for a fixed 
photon wave number $k=3 \pi /4a$. 
Consistent with our above interpretation, trapping is most
pronounced for zero detuning $ \delta = \Omega - \hbar \omega_{k}$
(recall that $\hbar \omega_{k=3\pi/4a} = \sqrt{2} J$).
Furthermore, proximity of the TLS resonance frequency to 
the band edge (or cut-off frequency of the waveguide) is 
clearly advantageous for realizing efficient trapping: For
frequencies near a band edge the multi-particle scattering
mechanism has to provide less additional energy for 
exciting the energetically closest bound state.
Less intuitive is the fact that there exists an optimal
coupling strength $V_{\rm opt} \sim J$ between TLS and 
waveguide modes for which maximal trapping occurs.
We have confirmed these findings for a number of different 
dispersion relations. For instance, we have extended  
Hamiltonian (\ref{eq:TB-U}) to include a next-nearest-neighbor 
hopping term $J^{(2)} \neq 0$ that allows us to significantly 
modify the cos-type dispersion relation of the tight-binding 
model (not shown). 
In addition, we have found analogous behavior for strictly 
linear dispersion relations with cut-off at finite energies
(not shown).

The above results suggest a certain robustness of the
trapping effect which we have further analyzed by 
qualitatively considering losses. This is accomplished
by coupling the TLS to a second waveguide that can 
de-excite the TLS into modes other than those of 
the original waveguide. 
The incorporation of this "loss channel" into the 
Hamiltonian (\ref{eq:TB-TLS}) thus proceeds by adding 
two additional terms analogous to, respectively, the 
first (hopping term $J^\prime$) and third term (coupling
term $V^\prime$) of the r.h.s. of (\ref{eq:TB-TLS}).
Clearly, the hopping term $J^\prime$ has to be chosen
such that the energy of the bound photon-atom states that 
are associated with the first waveguide and the TLS alone 
lies in the band of the second waveguide.
\begin{figure}
\includegraphics[width=0.45\textwidth]{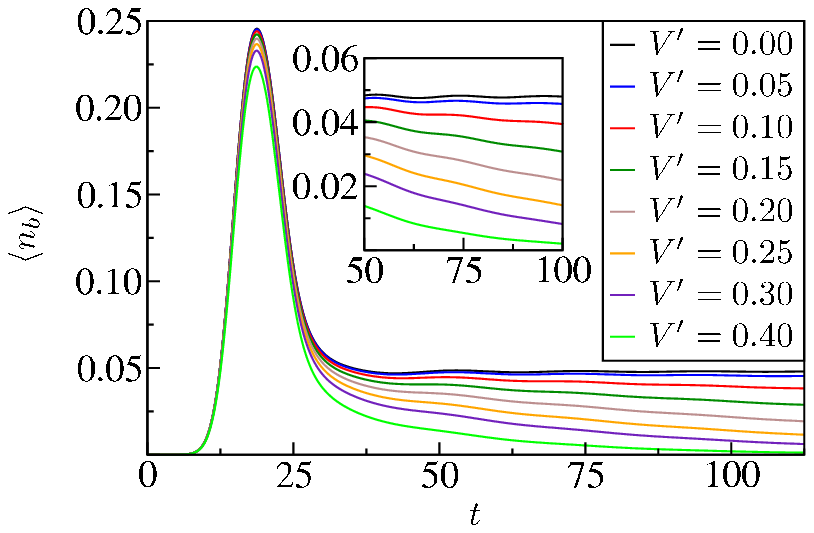}
\caption{\label{fig:stability}
(Color online)
Time evolution (in units of $\hbar / J$) of the impurity 
occupation $\langle n_b \rangle$ for initial two-photon 
states (see Fig. \ref{fig:bound-state-demo}) where a 
broad-band loss waveguide has been introduced. 
The corresponding system parameters are (loss waveguide's 
parameters are primed): 
$L = L^\prime = 399a$, $x_0 = x_0^\prime = 200a$, 
$\Omega = \sqrt{2}J$, $V=J$, and $J^\prime = 2J$. 
The strength of the coupling $V^\prime$ (in units of $J$) 
to the loss channel is varied.
The photon parameters are $x_c = 175a$, $s = 6a$, and
$k = 3 \pi / 4a$.
}
\end{figure}
In Fig. \ref{fig:stability}, we display the time evolution of
of the TLS' excited-state occupation for different coupling strengths
$V^\prime$ of the TLS to such a "broad-band loss waveguide". 
The trapping effect persists even for rather strong coupling 
to the loss channel. If we, for instance, interpret the 
coupling to the loss waveguide as a (admittedly crude) model 
for fabrication tolerances that in a quasi-one-dimensional
system couple strictly guided modes to a continuum of 
radiative modes, we are led to speculate that the trapping 
effect would be observable in experimentally accessible 
systems.

\begin{figure}
\includegraphics[width=0.45\textwidth]{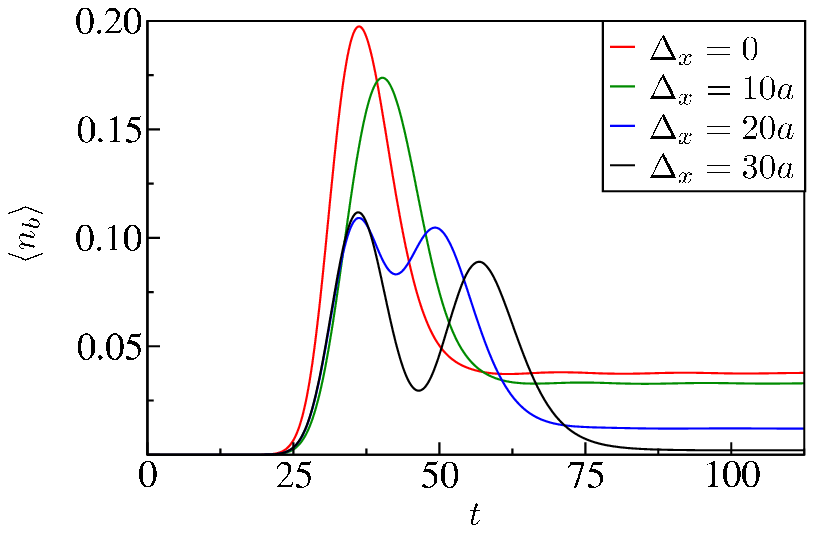}
\caption{\label{fig:control}
(Color online)
Time evolution (in units of $\hbar / J$) of the impurity occupation 
$\langle n_b \rangle$ for a system where two single-photon Gaussian
wave packets of width $s = 7a$ with different intial positions ($x_c^{(1)}$ 
and $x_c^{(2)}$) are launched from different sides towards the TLS. 
The corresponding system parameters are $L = 199a$, $x_0 = 100a$, 
$\Omega = \sqrt{2}J$, and $V = J$. The photon parameters are 
$x_c^{(1)} = 50a$, $k^{(1)} = 3 \pi / 4a = -k^{(2)}$, and the initial
position $x_c^{(2)} = 150a + \Delta_x$ is varied. 
}
\end{figure}

Finally, we have analyzed the possibility of tuning the 
trapped photon population at the TLS site. To do so, we have 
prepared two identical single-photon wave packets on different 
sides of the TLS and have launched them towards the TLS. By 
changing their initial relative separation to the TLS, we can
 exert some control over the multi-particle
processes in the Hamiltonian (\ref{eq:TB-U}) 
(see Fig. \ref{fig:control}). 
While the dynamcis is (expectedly) rather distinct for the 
different cases, we observe monotonic behavior of the 
trapped population: maximal trapping occurs for symmetrically 
launched pulses with zero relative initial distance. For
increased distances the trapped population decreases to 
zero once there is no overlap of the pulses at the TLS 
site. 

In conclusion, we have analyzed the dynamics of photon transport 
in waveguiding systems in the presence of a TLS within the context 
of a quantum many-body framework. Our reformulation (\ref{eq:TB-U}) 
allows us to identify strong multi-particle processes that may be 
utilized to excite and control photon-atom bound states. In turn,
this facilitates trapping of radiation at the TLS.
In addition, we have shown that this trapping effect exhibits 
a certain degree of robustness and can be found in a number of 
systems.
Since few-photon (or low intensity) coherent states are superpositions
of a few Fock states only (those that we have discussed in the present
work), we expect that the excitation and control of the photon-atom
bound states and associated effects will also occur in such situations.

Finally, we would like to emphasize the generality of our 
approach which is capable of treating systems with arbitrary dispersion 
relations and atom-field coupling strengths both in real and momentum
space.
Thus, the trapping of the photon population and its control suggest 
that such systems may be exploited for engineering photon entanglement 
as well as for the realization of quantum logic circuits in a number of 
systems that range from silicon integrated optical elements all the way 
to superconducting quantum circuits for microwave photons.

We acknowledge support by the Deutsche Forschungsgemeinschaft 
(DFG) and the State of Baden-W\"{u}rttemberg through the DFG-Center 
for Functional Nanostructures (CFN) within subprojects A1.2 and B2.10. 


\begin{thebibliography}{99}




\bibitem{Santori02}
C. Santori, D. Fattal, J. Vuckovic, G. S. Solomon, and Y. Yamamoto,
Nature {\bf 419}, 594 (2002).

\bibitem{Politi08}
A. Politi, M. J. Cryan, J. G. Rarity, S. Yu, and J. L. O'Brien,
Science {\bf 320}, 646 (2008).

\bibitem{Hofheinz08}
M. Hofheinz, E.M. Weig, M. Ansmann, R.C. Bialczak, E. Lucero et al.,
Nature {\bf 454}, 310 (2008).


\bibitem{Xia07}
F. Xia, L. Sekaric, and Y. Vlasov, 
Nature Photon. {\bf 1}, 65 (2007).

\bibitem{Notomi08}
M. Notomi, E. Kuramochi, and T. Tanabe,
Nature Photon. {\bf 2}, 741 (2008).


\bibitem{Fano}
U. Fano, Nuovo Cimento {\bf 12}, 156 (1935); U. Fano
Phys. Rev. {\bf 124}, 1866 (1961).

\bibitem{Shen05}
J. T. Shen and S. Fan,
Opt. Lett. {\bf 30}, 2001 (2005).

\bibitem{Zhou08}
L. Zhou, Z. R. Gong, Y. Liu, C. P. Sun, and F. Nori,
Phys. Rev. Lett. {\bf 101}, 100501 (2008).


\bibitem{Longo09}
P. Longo, P. Schmitteckert, and K. Busch,
J. Opt. A: Pure Appl. Opt. {\bf 11}, 114009 (2009).


\bibitem{Schmitteckert04}
P. Schmitteckert,
Phys. Rev. B {\bf 70}, 121302(R) (2004).

\bibitem{Ulbricht09}
T.~Ulbricht and P.~Schmitteckert
EPL {\bf 86}, No. 5,  57006 (2009).


\bibitem{Shen07a}
J. T. Shen and S. Fan,
Phys. Rev. Lett. {\bf 98}, 153003 (2007).

\bibitem{Shen07b}
J. T. Shen and S. Fan, 
Phys. Rev. A {\bf 76}, 062709 (2007).

\bibitem{Shi09}
T. Shi and C. P. Sun,
Phys. Rev. B {\bf 79}, 205111 (2009).

\bibitem{Dicke54}
R. H. Dicke,
Phys. Rev. {\bf 93}, 99 (1954).


\bibitem{Anderson61}
P. W. Anderson,
Phys. Rev. {\bf 124}, 41 (1961).

\bibitem{Quang94}
S. John and T. Quang,
Phys. Rev. A {\bf 50}, 1764 (1994).










%
%
%
%
%













\end{thebibliography}
\end{document}